\documentclass[oneside]{book}
\usepackage[T1]{fontenc}
\usepackage{a4wide}
\usepackage{graphics}
\usepackage{amssymb}

\makeatletter

\newcommand{\LyX}{L\kern-.1667em\lower.25em\hbox{Y}\kern-.125emX\spacefactor1000}

\makeatother

\begin{document}

\title{Study Notes on Numerical Solutions of the Wave Equation with the Finite Difference
Method\thanks{
Study supported by the PIBIC/CNPq undergraduate research program.
}}

\author{Artur B. Adib\thanks{
e-mail: adib@fisica.ufc.br 
}\\
Advisor: Carlos A. S. de Almeida\thanks{
e-mail: carlos@fisica.ufc.br
}}

\date{\emph{Departamento de Física, Universidade Federal do Ceará, Brazil}}

\maketitle

\chapter*{Preface}

The simulation of the dynamics of classical field theories is currently gaining
some attention from the high-energy community, mainly in the context of statistical
field theory. Recent papers show that, in some particular but important conditions,
\emph{classical} field theories are a very good approximation to the \emph{quantum}
evolution of fields at finite temperature (see, for instance, \cite{gleiser, aarts}).
Also, in the context of condensed-matter systems, the dynamics of effective
classical fields has proven to be a very efficient tool in describing both the
equilibrium and non-equilibrium properties of systems such as ferromagnets \cite{boyanovsky}.
In order to simulate these theories, they must be first discretized and cast
on a lattice, a job far from trivial to do in a consistent manner \cite{borrill}.
After the discretization, the differential equations of motion transform themselves
into \emph{finite difference} equations. Before doing any sort of useful calculation
in the physical context above, one is supposed to understand the basic foundations
of the numerical method, the study of which is the objective of this monograph.

Following this motivation, I will present, in an introductory way, the Finite
Difference method for hyperbolic equations, focusing on a method which has second
order precision both in time and space (the so-called \emph{staggered leapfrog}
method) and applying it to the case of the 1d and 2d wave equation. A brief
derivation of the energy and equation of motion of a wave is done before the
numerical part in order to make the transition from the continuum to the lattice
clearer. 

To illustrate the extension of the method to more complex equations, I also
add dissipative terms of the kind \( -\eta \dot{u} \) into the equations. I
also briefly discuss the \emph{von Neumann} numerical stability analysis and
the \emph{Courant} criterion, two of the most popular in the literature. In
the end I present some numerical results obtained with the leapfrog algorithm,
illustrating the importance of the lattice resolution through energy plots. 

I have tried to collect, in a concise way, the main steps necessary to have
a stable algorithm to solve wave-like equations. More sophisticated versions
of these equations should be handled with care, and accompanied of a rigorous
study of convergence and stability which could be found in the references cited
in the end of this work.

\tableofcontents

\chapter{The Wave Equation}

\section{Introduction\label{sec: eqonda_intro}}

Partial Differential Equations (from now on simply \emph{PDEs}) are divided
in the literature basically in three kinds: parabolic, elliptic and hyperbolic
(the criterion of classification of these equations can be found in \cite[Chap. 8]{arfken}).
In this work we will be interested mainly on hyperbolic equations, of which
the wave equation is the paradigm:

\begin{equation}
\label{eqonda}
\nabla ^{2}u=\frac{1}{v^{2}}\frac{\partial ^{2}u}{\partial t^{2}},
\end{equation}
where \( v^{2}=\frac{\tau }{\lambda } \) is the square of the wave velocity
in the medium, which in the case of a free string could be determined by the
tension \( \tau  \) and the mass density per length unit \( \lambda  \).

From the strict numerical point of view, the distinction between these classes
of PDEs isn't of much importance \cite{recipes}. There is, however, another
sort of classification of PDEs which is relevant for numerical purposes: the
\emph{initial value} \emph{problems} (which include the case of the hyperbolic
equations) and the \emph{boundary condition problems} (which include, for instance,
parabolic equations)\emph{.} In this work we will restrict ourselves to initial
value problems. See reference \cite{recipes} for a good introduction to boundary
condition problems.

In the equation (\ref{eqonda}) we could still add a dissipative term proportional
to the first power of the time derivative of \( u \), i.e.,
\begin{equation}
\label{eqonda_dissp}
\tau \nabla ^{2}u=\lambda \frac{\partial ^{2}u}{\partial t^{2}}+\eta \frac{\partial u}{\partial t},
\end{equation}
where \( \eta  \) is the viscosity coefficient.

Our first step will be to derive the wave equation from a simple mechanical
analysis of the free rope. Being this a well known problem in classical mechanics,
we will go through only the main steps of it (for a more complete treatment
of the problem of the free string, see, for example, \cite[Chaps. 8 and 9]{symon}).
Once we are done with the 1-d wave equation, we will proceed further to the
2-d case, which isn't as abundant in the literature as the 1-d case.

\section{Waves in 1-dimension (the free string) \label{sec: eqonda_1d}}

\subsection{Equation of Motion}

\begin{figure}
{\centering \resizebox*{0.4\textwidth}{!}{\includegraphics{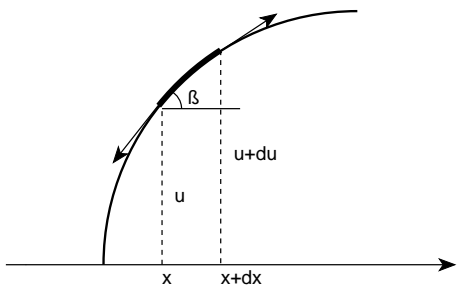}} \par}

\caption{\label{fig: corda}Representing the tension forces acting on an infinitesimal
element of the rope.}
\end{figure}

Figure \ref{fig: corda} gives us an idea of a mass element \( dm \) with linear
dimension \( dx \) subject to tension forces. We are interested on the vertical
displacement of this mass element, so, for this direction, we could write the
resulting force:
\begin{equation}
\label{dFu_resultante}
dF_{u}=\left. \vec{\tau }\cdot \hat{u}\right| _{x+dx}-\left. \vec{\tau }\cdot \hat{u}\right| _{x},
\end{equation}
where \( \vec{\tau } \) is the tension and the unit vector \( \hat{u} \) refers
to the vertical direction. Within the domain of smooth deformations of the string
(i.e., small \( \beta  \)), we could write:
\begin{equation}
\label{peq_osc_1d}
\tau _{u}\equiv \vec{\tau }\cdot \hat{u}=\tau \sin \beta \approx \tau \tan \beta =\tau \frac{\partial u}{\partial x}
\end{equation}

We notice now that (\ref{dFu_resultante}) could be written as:
\[
dF_{u}=\frac{\left. \tau _{u}\right| _{x+dx}-\left. \tau _{u}\right| _{x}}{dx}dx=\frac{\partial \tau _{u}}{\partial x}dx=\frac{\partial }{\partial x}\left( \tau \frac{\partial u}{\partial x}\right) dx\]

We will now restrict ourselves to the case of constant tensions along the rope,
so that:
\begin{equation}
\label{dFu_em_dudx}
dF_{u}=\tau \frac{\partial ^{2}u}{\partial x^{2}}dx
\end{equation}

Equating this with Newton's second law
\[
dF_{u}=dm\frac{\partial ^{2}u}{\partial t^{2}}=\lambda \frac{\partial ^{2}u}{\partial t^{2}}dx,\]
where \( \lambda  \) is the linear mass density, we obtain then the wave equation
for a free string:
\begin{equation}
\label{eqonda_corda}
\tau \frac{\partial ^{2}u}{\partial x^{2}}=\lambda \frac{\partial ^{2}u}{\partial t^{2}}
\end{equation}

\subsection{Energy}

The kinetic energy could be evaluated in a straightforward manner, integrating
the kinetic energy term for a representative mass element:
\[
dT=\frac{1}{2}dm\cdot v_{u}^{2},\]
with \( dm=\lambda \cdot dx \) and \( v_{u}=\frac{\partial u}{\partial t} \),
in other words,
\begin{equation}
\label{T_corda}
T=\int dT=\frac{\lambda }{2}\int _{0}^{l}\left( \frac{\partial u}{\partial t}\right) ^{2}dx
\end{equation}

The potential energy can be obtained by calculating the work necessary to bring
the string from a ``trivial'' configuration \( u(x,0)=0 \) to the configuration
at which we want to evaluate the potential energy \( u(x,t) \). We will fix
the boundary conditions \( u(0,t)=u(l,t)=0 \) (string tied at the ends) and
as a consequence of this, \( \partial _{t}u(0,t)=\partial _{t}u(l,t)=0 \).
The potential energy relative to the work necessary to change of \( \delta u \)
the configuration of an element of the string in an interval of time \( dt \)
is:
\[
\delta V=-dF_{u}\cdot \delta u=-dF_{u}\cdot \left( \frac{\partial u}{\partial t}\right) dt\]

Therefore, the potential energy of the \emph{whole} string in this \emph{same
interval of time} is:
\[
dV=-dt\cdot \int dF_{u}\left( \frac{\partial u}{\partial t}\right) \]

Substituting (\ref{dFu_em_dudx}) in the latter we get:
\[
dV=-dt\cdot \int _{0}^{l}\tau \left( \frac{\partial ^{2}u}{\partial x^{2}}\right) \left( \frac{\partial u}{\partial t}\right) dx\]

We are however interested on \( V\left[ u(x,t)\right]  \), so integrating in
time and using the boundary conditions above, we have:
\begin{eqnarray*}
V & = & \int dV=-\tau \int _{0}^{t}dt\int _{0}^{l}dx\left( \frac{\partial u}{\partial t}\right) \left( \frac{\partial ^{2}u}{\partial x^{2}}\right) =\\
 & = & -\tau \int _{0}^{t}dt\left\{ \left. \frac{\partial u}{\partial t}\frac{\partial u}{\partial x}\right| _{0}^{l}-\int _{0}^{l}\frac{\partial u}{\partial x}\frac{\partial ^{2}u}{\partial x\partial t}dx\right\} =\\
 & = & \tau \int _{0}^{t}dt\int _{0}^{l}\frac{\partial u}{\partial x}\frac{\partial ^{2}u}{\partial x\partial t}dx=\\
 & = & \tau \int _{0}^{t}dt\frac{1}{2}\frac{\partial }{\partial t}\int _{0}^{l}\left( \frac{\partial u}{\partial x}\right) ^{2}dx=\\
 & = & \frac{\tau }{2}\left. \int _{0}^{l}\left( \frac{\partial u}{\partial x}\right) ^{2}dx\right| _{0}^{t}\Rightarrow 
\end{eqnarray*}

\begin{equation}
\label{V_corda}
V=\frac{\tau }{2}\int _{0}^{l}\left( \frac{\partial u}{\partial x}\right) ^{2}dx
\end{equation}

With (\ref{T_corda}) and (\ref{V_corda}) we have finally the total energy
of the rope:
\begin{equation}
\label{E_corda_lamb_tau}
E=\frac{\lambda }{2}\int _{0}^{l}\left( \frac{\partial u}{\partial t}\right) ^{2}dx+\frac{\tau }{2}\int _{0}^{l}\left( \frac{\partial u}{\partial x}\right) ^{2}dx
\end{equation}

In our applications we will take \( \lambda =\tau  \) such that \( v^{2}=1 \)
and (\ref{E_corda_lamb_tau}) assumes the simple form:
\begin{equation}
\label{E_corda}
E=\frac{1}{2}\int _{0}^{l}\left[ \left( \frac{\partial u}{\partial t}\right) ^{2}+\left( \frac{\partial u}{\partial x}\right) ^{2}\right] dx
\end{equation}

This energy equation will be very useful to test our algorithms through an analysis
of conservation (or dissipation) during the dynamical evolution of the system.

\section{Waves in 2-dimensions (the free membrane)\label{sec: eqonda_2d}}

\subsection{Equation of Motion}

\begin{figure}
{\centering \resizebox*{0.4\textwidth}{!}{\includegraphics{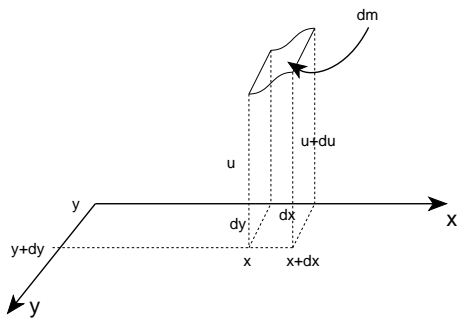}} \par}

\caption{\label{fig: elemento_dxdy}Representing a mass element \protect\( dm\protect \)
of dimensions \protect\( dxdy\protect \). The element is subject to tension
forces on each side (analogous to the borders of the mass element in the 1-d
case), being these forces orthogonal to the axes of the sides. Only the variation
with respect to \protect\( x\protect \) of \protect\( u\protect \) is drawn
(i.e., \protect\( u\protect \) and \protect\( u+du\protect \) in the figure
are displacements of \protect\( u(x,y)\protect \) keeping \protect\( y\protect \)
constant and varying \protect\( x\protect \)).}
\end{figure}

In its two dimensional version, the wave equation could be describing a membrane,
a liquid surface, or some ``coarse-grained'' field in the surface physics,
to cite a few. In the case of the membrane or other elastic surface, the oscillations
are also constrained to be small (analogously to the 1-d string). 

We notice now that the additional dimension forces us to define the tension
``per unit length'':
\begin{equation}
\label{f_unid_comp}
f=\frac{\tau }{l}
\end{equation}

This force per unit length could be understood with a simple example: stretch
a tape of width \( l \) from its extremities with force \( \tau  \). We can't
ask the force on a \emph{point} of the tape, but only on some element of some
definite length and width (of course, this element could be differential, playing
the same role of a linear differential element in the case of the string). Formula
(\ref{f_unid_comp}), times the length of the element, then gives you the resulting
force (tension) on the element. In this way, we extend the equation (\ref{dFu_resultante})
to two dimensions:
\begin{equation}
\label{dFu_em_dxdy}
dF_{u}=\left[ \left. \vec{f}_{x}\cdot \hat{u}\right| _{x+dx,y}dy-\left. \vec{f}_{x}\cdot \hat{u}\right| _{x,y}dy\right] +\left[ \left. \vec{f}_{y}\cdot \hat{u}\right| _{x,y+dy}dx-\left. \vec{f}_{y}\cdot \hat{u}\right| _{x,y}dx\right] ,
\end{equation}
where \( \left. \vec{f}_{x}\cdot \hat{u}\right| _{x,y}dy\equiv \left. f_{x,u}\right| _{x,y}dy \)
is the \( x \) component of the tension in the direction \( \hat{u} \) acting
on the side defined by the points \( (x,y) \) e \( (x,y+dy) \), and so on.
We are going to suppose that the forces on the sides of the elements are orthogonal
to their axes, which is the same as decomposing the tension force on \( dm \)
into four components, one for each side (notice, however, that we have effectively
only \emph{two} resulting components, to wit \emph{\( \hat{x} \)} and \( \hat{y} \)).
Doing this we won't need to emphasize the tension components along \( x \)
or \( y \), \( \left. \vec{f}_{x}\cdot \hat{u}\right| _{x,y} \) becoming simply
\( \left. f_{u}\right| _{x,y} \) and so on. Nevertheless it \emph{is} still
important, for what we said above, to know \emph{what} side we are talking about.
So, in the regime of small vibrations (i.e., small angles of deformation), we
could find \( f_{u} \) analogously to the string case
\begin{eqnarray}
f_{u}dy & = & f\frac{\partial u}{\partial x}dy\\
f_{u}dx & = & f\frac{\partial u}{\partial y}dx,\label{peq_osc_2d} 
\end{eqnarray}
where \( f_{u}dy \) and \( f_{u}dx \) are the tensions in the direction \( \hat{u} \)
on a side of length \( dy \) and \( dx \) along \( y \) and \( x \), respectively.
We emphasize that, with this notation plus the knowledge of the \emph{point}
where we are going to evaluate the derivatives, we have a complete specification
of the side on which the tension acts\footnote{
Indeed, once specified the \emph{beginning} of the side with the pair \( (x,y) \),
specifying the length with \( dx \) or \( dy \) furnishes us with the \emph{direction}
of the side in question. This is sufficient to localize it, since the \( z \)
\emph{}coordinate is unambiguously determined via \( z=u(x,y) \).
}. With all this in hands, Eq. (\ref{dFu_em_dxdy}) becomes:
\[
dF_{u}=\left[ \left. f_{u}\right| _{x+dx,y}dy-\left. f_{u}\right| _{x,y}dy\right] +\left[ \left. f_{u}\right| _{x,y+dy}dx-\left. f_{u}\right| _{x,y}dx\right] \]

With calculations analogous to those of the previous section, we have:
\begin{eqnarray}
dF_{u} & = & \frac{\left. f_{u}\right| _{x+dx,y}-\left. f_{u}\right| _{x,y}}{dx}dxdy+\frac{\left. f_{u}\right| _{x,y+dy}-\left. f_{u}\right| _{x,y}}{dy}dydx=\\
 & = & \frac{\partial f_{u}}{\partial x}dxdy+\frac{\partial f_{u}}{\partial y}dxdy=\\
 & = & f\frac{\partial ^{2}u}{\partial x^{2}}dxdy+f\frac{\partial ^{2}u}{\partial y^{2}}dxdy\label{dFu_result_dxdy} 
\end{eqnarray}

With Newton's second law we obtain:
\begin{eqnarray*}
f\frac{\partial ^{2}u}{\partial x^{2}}dxdy+f\frac{\partial ^{2}u}{\partial y^{2}}dxdy & = & dm\frac{\partial ^{2}u}{\partial t^{2}}\Rightarrow \\
f\frac{\partial ^{2}u}{\partial x^{2}}dxdy+f\frac{\partial ^{2}u}{\partial y^{2}}dxdy & = & \sigma dxdy\frac{\partial ^{2}u}{\partial t^{2}}\Rightarrow \\
f\left( \frac{\partial ^{2}u}{\partial x^{2}}+\frac{\partial ^{2}u}{\partial y^{2}}\right)  & = & \sigma \frac{\partial ^{2}u}{\partial t^{2}}\Rightarrow 
\end{eqnarray*}

\begin{equation}
\label{eqonda_2d}
\frac{\partial ^{2}u}{\partial x^{2}}+\frac{\partial ^{2}u}{\partial y^{2}}=\frac{1}{v^{2}}\frac{\partial ^{2}u}{\partial t^{2}},
\end{equation}
which is the desired wave equation for two dimensions, with \( v^{2}=\frac{f}{\sigma } \)
and \( \sigma  \) the surface mass density.

\subsection{Energy}

The derivation of the total energy is done in the same manner as the 1d case.
We will consider a surface \( z=u(x,y,t) \) with support of dimension \( l\times l \),
subject to the boundary conditions \( \left. u\right| _{boundary}\equiv u(0,y,t)=u(l,y,t)=u(x,0,t)=u(x,l,t)=0 \)
and \( \left. \dot{u}\right| _{boundary}\equiv \dot{u}(0,y,t)=\dot{u}(l,y,t)=\dot{u}(x,0,t)=\dot{u}(x,l,t)=0 \),
where \( \dot{u}\equiv \partial u/\partial t \). Let us begin with the kinetic
term:
\begin{eqnarray}
dT & = & \frac{1}{2}dm\cdot \dot{u}^{2}=\frac{\sigma }{2}\dot{u}^{2}dxdy\Rightarrow \\
T & = & \frac{\sigma }{2}\int _{0}^{l}\int _{0}^{l}\left( \frac{\partial u}{\partial t}\right) ^{2}dxdy,\label{T_2d} 
\end{eqnarray}
where \( \sigma  \) is the surface mass density.

The potential energy is obtained in an analogous way to the Section (\ref{sec: eqonda_1d}):
\begin{eqnarray*}
dV & = & -dt\cdot \int dF_{u}\left( \frac{\partial u}{\partial t}\right) =\\
 & = & -dt\cdot \int _{0}^{l}\int _{0}^{l}f\left( \frac{\partial ^{2}u}{\partial x^{2}}+\frac{\partial ^{2}u}{\partial y^{2}}\right) \left( \frac{\partial u}{\partial t}\right) dxdy\Rightarrow \\
V & = & -f\int _{0}^{t}dt\left\{ \int _{0}^{l}dy\int _{0}^{l}dx\left( \frac{\partial ^{2}u}{\partial x^{2}}\frac{\partial u}{\partial t}\right) +\int _{0}^{l}dx\int _{0}^{l}dy\left( \frac{\partial ^{2}u}{\partial y^{2}}\frac{\partial u}{\partial t}\right) \right\} =\\
 & = & f\int _{0}^{t}dt\left\{ \int _{0}^{l}dy\frac{1}{2}\frac{\partial }{\partial t}\int _{0}^{l}\left( \frac{\partial u}{\partial x}\right) ^{2}dx+\int _{0}^{l}dx\frac{1}{2}\frac{\partial }{\partial t}\int _{0}^{l}\left( \frac{\partial u}{\partial y}\right) ^{2}dy\right\} =\\
 & = & \frac{f}{2}\int _{0}^{t}dt\frac{\partial }{\partial t}\left\{ \int _{0}^{l}\int _{0}^{l}\left[ \left( \frac{\partial u}{\partial x}\right) ^{2}+\left( \frac{\partial u}{\partial y}\right) ^{2}\right] dxdy\right\} =\\
 & = & \frac{f}{2}\left. \int _{0}^{l}\int _{0}^{l}\left[ \left( \frac{\partial u}{\partial x}\right) ^{2}+\left( \frac{\partial u}{\partial y}\right) ^{2}\right] dxdy\right| _{0}^{t}\Rightarrow 
\end{eqnarray*}

\begin{equation}
\label{V_2d}
V=\frac{f}{2}\int _{0}^{l}\int _{0}^{l}\left[ \left( \frac{\partial u}{\partial x}\right) ^{2}+\left( \frac{\partial u}{\partial y}\right) ^{2}\right] dxdy
\end{equation}
 Then the total energy for our usual condition \( f=\sigma \Rightarrow v^{2}=1 \)
is:
\[
E=\frac{1}{2}\int _{0}^{l}\int _{0}^{l}\left[ \left( \frac{\partial u}{\partial x}\right) ^{2}+\left( \frac{\partial u}{\partial y}\right) ^{2}+\left( \frac{\partial u}{\partial t}\right) ^{2}\right] dxdy\]
or, in a more general way (we are going to take as granted the result for more
than two dimensions),
\[
E=\int d^{n}\vec{r}\left[ \frac{1}{2}(\vec{\nabla }u)^{2}+\frac{1}{2}\dot{u}^{2}\right] \]

Once again I emphasize that these results are important for the verification
of the stability of our numerical analysis. This derivation is shown here not
only as an exercise, but also because I couldn't find the 2-d version in any
textbook.

\chapter{Finite Differences}

\section{Introduction}

Differently from the non-approximate analytical solutions of PDEs in the continuum
(for instance, those obtained through variable separation and subsequent integration),
numerical solutions obtained in a computer have limited precision\footnote{
In this point it is worth mentioning that there are basically two ways of solving
a mathematical problem with the aid of a computer: symbolically and numerically.
Symbolic methods deal fundamentally with algebraic manipulations and do not
involve explicit numerical calculations, giving us an analytical form (whenever
possible) to the desired problem. It is, however, widely understood that non-linear
theories hardly have a closed-form solution, and even if they do, it is often
a lot complicated and requires an understanding of very sophisticated tools.
Whenever this is the case, one often resorts to the numerical approach, which
doesn't furnish us with an analytical closed-form solution, but could give very
precise numerical estimates for the solution of the problem. It has been used
since the very beggining of the computer era, and today it is sometimes the
\emph{only} tool people have to attack some problems, pervading its use in almost
every discipline of science and technology.
}. It is due to the way in which computers store data and also because of their
limited memory. After all, how could we write in decimal notation (or in any
other base) an irrational number like \( \sqrt{2} \) making use of a finite
number of digits? In this work we won't stick with rigorous derivations of the
theorems nor of most of the results presented. The references listed in the
end should be considered for this end.

The central idea of numerical methods is quite simple: to give finite precision
(``the discrete'') to those objects endowed with infinite precision (``the
continuum''). By \emph{discretize} we understand to transform continuum variables
like \( x,y,..,z \) into a set of discrete values \( \{x_{i}\},\{y_{i}\},...,\{z_{i}\} \),
where \( i \) runs over a finite number of values, thus sampling the wholeness
of the original variables. As a consequence of this discretization, integrals
become sums and derivatives turns out to mere differences of finite quantities
(hence the name ``finite differences''). I illustrate below these ideas:
\begin{eqnarray}
\int f(x)dx & = & \lim _{\delta x\rightarrow 0}\sum _{n}f(n\delta x)\delta x\, \rightarrow \, \sum _{n}f(n\Delta x)\Delta x\\
\frac{df(x)}{dx} & = & \lim _{\delta x\rightarrow 0}\frac{f(x+\delta x)-f(x)}{\delta x}\, \rightarrow \, \frac{f(x+\Delta x)-f(x)}{\Delta x},
\end{eqnarray}
where \( \delta x \) is a variable with \emph{infinite} precision (thus its
value could be as small as we want) and \( \Delta x\ll 1 \) is a variable with
\emph{finite} precision, which under the computational point of view is the
limiting case analogous to \( \delta x \). We could naively expect that, the
smaller the value of \( \Delta x \), the closer we are to the continuum theory.
This would be indeed true if computers didn't have finite precision! The closer
your significant digits get to the limiting precision of the computer, the \emph{worse}
is your approximation, because it will introduce the well-known ``round-off
errors'', which are basically truncation errors. The reference \cite{cheney}
has a somewhat lengthy discussion about computational issues like this.

\section{Difference Equations}

\emph{Difference} equations are to a computer in the same way as \emph{differential}
equations are to a good mathematician. That is, if you have a problem in the
form of a differential equation, the most straightforward way of solving it
is to transform your derivatives into differences, so that you finish with \emph{}an
\emph{algebraic} difference equation. This turns out to be necessary for what
we said about the limitations of a computer\footnote{
There are, however, more sophisticated methods like Finite Elements, but the
fact that one needs to get rid of differentials transcends these methods when
we are talking about numerical solutions.
}.

As a trivial example, take the ordinary differential equation:
\begin{equation}
\label{ode}
\frac{df}{dx}=g(x)
\end{equation}

Using a first-order Taylor expansion (see Appendix \ref{sec: Ap_taylor}) for
\( f(x) \),
\[
f(x+h)\approx f(x)+f'(x)h\Rightarrow \]
 
\[
f'(x)\approx \frac{f(x+h)-f(x)}{h}\]
 we obtain the \emph{Euler form} for the Eq. (\ref{ode}):
\[
\frac{f(x+h)-f(x)}{h}\approx g(x)\]

Notice that this equation involves only differences as we said above, and to
solve it in a computer we shall need the following \emph{iterative relation}
obtained directly from the above equation:
\[
f(x+h)=hg(x)+f(x)\]
or, in the traditional numerical notation:
\begin{equation}
\label{Euler_discreto}
f_{n+1}=hg_{n}+f_{n}
\end{equation}

Technically, once provided both the initial condition (for instance, \( f_{0}=0 \))
and the functional form of \( g_{n}=g(x_{n}) \), we could solve Eq. (\ref{Euler_discreto})
by iterating it in a program loop.

In spite of its simple form, the Euler method is far from being useful for realistic
equations; it could give rise to a completely erroneous approximation. Higher
order expansions are frequently used in order to obtain equations with reduced
error (see again Appendix \ref{sec: Ap_taylor} for some of these expansions).
However, these higher order approximations are \emph{also} subject to serious
problems, like the lack of stability or convergence, so the problem is ubiquitous
and has been one of the most attacked problems in the so-called ``numerical
analysis'', a relatively modern branch of mathematics. I will say a little
bit more later about these issues on convergence and stability.

It has already been said that the relevance of the classification of PDEs lies
in their ``nature''; those of initial value have a completely different way
of solving numerically from those of boundary values. The latter kind doesn't
evolve in time. It's the classic case of the Poisson equation which could be
describing a thermostatic or an electrostatic system:
\begin{equation}
\label{poisson}
\frac{\partial ^{2}u(x,y)}{\partial x^{2}}+\frac{\partial ^{2}u(x,y)}{\partial y^{2}}=f(x,y)
\end{equation}

Using the expansions from Appendix \ref{sec: Ap_taylor}, we have: 
\begin{equation}
\label{poisson_discreta}
\frac{u_{i+1,j}-2u_{i,j}+u_{i-1,j}}{h^{2}}+\frac{u_{i,j+1}-2u_{i,j}+u_{i,j-1}}{h^{2}}=f_{i,j},
\end{equation}
where we took \( h \) as the \emph{lattice} \emph{spacing} (also grid or net
resolution) both for the coordinate \( x \) and \( y \) (\( x\rightarrow x_{n}=nh \)
and \( y\rightarrow y_{n}=nh \)). The indices of this equation then correspond
to \emph{sites} in this lattice (a.k.a. lattice points), and sweep from \( 0 \)
to the number of sites \( N_{i} \) or \( N_{j} \). The problem becomes then
to solve the equations given by (\ref{poisson_discreta}) simultaneously for
the various \( u_{i,j} \). There are very interesting methods to solve this
sort of problem which could be found in the references \cite{recipes, cheney}.

Our work, however, is directed towards initial value problems which, as the
very name suggests, deal with temporal evolutions starting from certain ``initial
values'' at the ``zero'' instant. It is the typical case of the wave equation
already presented, or of the diffusion equation

\begin{equation}
\label{difusao}
\frac{\partial ^{2}u}{\partial x^{2}}=\frac{\partial u}{\partial t}.
\end{equation}

Our focus goes even finer, since we will deal only with ``explicit discretizations'',
which could be understood as those which could be solved iteratively, that is,
we could solve the difference equation for \( u(x,y,t+\Delta t) \) explicitly
in terms of the other variables \( u(x',y',t) \) at the instant \( t \) (for
instance, Euler's equation above is an explicit method). Implicit methods need
a different approach, which often involves the solution of linear systems by
using matrices (again \cite{recipes, cheney} do very well in these matters).

\section{The von Neumann Stability Analysis}

How should we know, after transforming a differential equation into finite differences,
if the calculated solution is a stable one? By \emph{numerically stable solutions}
we understand those in which the error \( z_{m}^{n} \) between the correct
theoretical solution \( u(x_{m},t_{n}) \) and the numerical solution \( U_{m}^{n} \)
does not diverge (i.e., is limited) as \( n\rightarrow \infty  \) (\( t\rightarrow \infty  \)),
in other words: 
\begin{equation}
\label{erro_num}
z_{m}^{n}\equiv u(x_{m},t_{n})-U_{m}^{n}<\epsilon ,
\end{equation}
for any \( n \), where the lower indices are spatial and the upper ones are
temporal, and \( \epsilon  \) is a \emph{finite} real value. For instance,
an unstable discretization describing a vibrating string could be easily detected
watching the energy of the system for a while: a divergent energy would certainly
arise. Fortunately, there is a useful tool to identify unstable finite difference
equations prior to simulating it, known as \emph{von Neumann stability analysis}
\cite{recipes, mitchell}, which could be applied to a difference equation to
preview its numerical behavior. 

The von Neumann method consists essentially in expanding the numerical error
\( z_{m}^{n} \) in a discrete harmonic Fourier series:
\begin{equation}
\label{decomp_neumann}
z_{m}^{n}=\sum _{r}a_{r}(t_{n})e^{ik_{r}x_{m}}
\end{equation}
and analyzing if \( a_{r}(t_{n}) \) increases (or decreases) as \( t\rightarrow \infty  \)
(technically, if \( a_{r}(t) \) decreases when \( t\rightarrow \infty  \)
we have a \emph{numerical dissipation}, which is usually harmless). It is then
easy to see that if \( a_{r}(t_{n}) \) isn't divergent for any \( n \) and
\( m \) we will have a stable solution. This analysis is somewhat simple, since
it is sufficient to study the behavior of a \emph{single general} term of the
series, for if we prove that this general term of the series could have a certain
pathological behavior (like diverging for \( n\rightarrow \infty  \)), then
the \emph{whole} solution is compromised; otherwise, our solution is stable.

Mitchell and Griffiths \cite{mitchell} show that \( z_{m}^{n} \) given by
(\ref{erro_num}) satisfy the very same difference equation for \( u_{m}^{n} \).
Hence, if we take a certain \( z_{m}^{n} \) such that \( \left| z_{m}^{0}\right| =1 \)
and put it into the difference equation, we could achieve the desired stability
condition. One possible \( z_{m}^{n} \) satisfying the criteria above is:
\begin{equation}
\label{zmn}
z_{m}^{n}=e^{\alpha n\Delta t}e^{i\beta m\Delta x}
\end{equation}

Indeed, notice that for \( n=0 \) we have \( \left| z_{m}^{0}\right| =1 \),
and with \( \alpha  \) and \( \beta  \) arbitrary values we satisfy the above
discussion. With this expression, we could now write the stability condition
for the von Neumann analysis:
\begin{equation}
\label{cond_neumann}
\left| \xi ^{n}\right| \leq 1,
\end{equation}
where \( \xi =e^{\alpha \Delta t} \) is the \emph{amplification factor}\textbf{.}
In summary, putting the error given by
\begin{equation}
\label{zmn_neumann}
z_{m}^{n}=\xi ^{n}e^{i\beta m\Delta x}
\end{equation}
into the difference equation, together with (\ref{cond_neumann}), we get the
necessary condition for stability.

\section{The Courant Condition}

\begin{figure}
{\centering \resizebox*{0.4\textwidth}{!}{\includegraphics{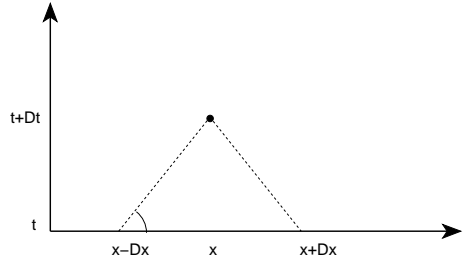}} \par}

\caption{\label{fig: courant}Representing an integration algorithm that needs the values
\protect\( u(x-Dx,t)\protect \) and \protect\( u(x+Dx,t)\protect \) to obtain
\protect\( u(x,t+Dt)\protect \). The ratio \protect\( Dx/Dt\protect \) is
then the ``maximum speed'' with which an information in the algorithm could
propagate.}
\end{figure}

Another important condition that we should pay some attention in initial value
problems is related to the \emph{speed} with which information could propagate
in the difference equation. We could visualize the problem in the scheme of
Figure \ref{fig: courant}.

It could be shown \cite{recipes} that, applying von Neumann's condition for
hyperbolic problems we arrive at the \emph{Courant condition}: if the ``physical''
wave velocity \( |v| \) in a differential equation is greater than the ``algorithm
speed'' \( \Delta x/\Delta t \), then the scheme is unstable. We have therefore
the following expression for the Courant condition: 
\begin{equation}
\label{cond_Courant}
\left| v\right| \leq \frac{\Delta x}{\Delta t}
\end{equation}

\section{The Staggered Leapfrog Algorithm}

In this section we present the basic idea behind the staggered leapfrog method,
which is a very efficient second-order integration scheme without numerical
dissipation. 

The method is for differential equations which could be cast on the form of
a flux-conservative one:

\begin{equation}
\label{flux-cons}
\frac{\partial \vec{u}}{\partial t}=-\frac{\partial \vec{F}}{\partial x}
\end{equation}
For instance, the wave equation could be written in this form when we make the
following substitutions (to ease the notation, from now on we will take \( v=\tau =\lambda =\sigma =1 \)):

\[
\frac{\partial }{\partial t}\frac{\partial u}{\partial t}=\frac{\partial }{\partial x}\frac{\partial u}{\partial x}\Rightarrow \]

\[
\frac{\partial s}{\partial t}=\frac{\partial r}{\partial x},\: \: \: \: \frac{\partial s}{\partial x}=\frac{\partial r}{\partial t}\]
where \( s=\partial _{t}u \) and \( r=\partial _{x}u \). So, putting in the
required form, we have:

\[
\frac{\partial }{\partial t}\left[ \begin{array}{c}
s\\
r
\end{array}\right] =-\frac{\partial }{\partial x}\left[ \begin{array}{c}
-r\\
-s
\end{array}\right] \]

The name ``leapfrog'' refers to the fact that both the time and spatial derivatives
are to be evaluated using a centered difference scheme, so the integration is
done from time-step \( n-1 \) to \( n+1, \) ``leaping'' over the spatial
derivatives which are evaluated at step \( n \). The centered difference is
a second-order one, giving for the flux-conservative equation:

\[
\frac{\vec{u}_{i}^{n+1}-\vec{u}_{i}^{n-1}}{2\Delta t}=\frac{\vec{F}_{i+1}^{n}-\vec{F}_{i-1}^{n}}{2\Delta x}\]

In the case of second-order differential equations, we need also the first derivative
in order to integrate for the next step. In the present scheme, these derivatives
are evaluated at some ``artificial'' points between the ``real'' ones, and
hence are said to be ``staggered'' with regard to the variable \( u^{n} \)
which is evaluated at the real points. Take for example the case below in which
an arbitrary function \( f[u] \) does not depend on any time derivative of
\( u \):

\begin{equation}
\label{leap-geral}
\frac{\partial s(x,t)}{\partial t}=f[u(x,t)],
\end{equation}
where \( s=\partial _{t}u \) and, for the wave equation, \( f(u)=\partial _{x}^{2}u \).
According to the above discussion, the derivative \( s \) is to be evaluated
at points \( n+1/2,n+3/2,... \), so we have\footnote{
Notice that the centered differences are now evaluated with a step of \( \Delta t/2 \).
}:

\[
s_{i}^{n+1/2}=\left. \frac{\partial u}{\partial t}\right| _{i}^{n+1/2}\approx \frac{u_{i}^{n+1}-u_{i}^{n}}{\Delta t}\Rightarrow \]

\begin{equation}
\label{leap-unp1}
u_{i}^{n+1}=u_{i}^{n}+\Delta ts_{i}^{n+1/2}
\end{equation}
But Eq. (\ref{leap-geral}) furnishes us with an expression for \( s_{i}^{n+1/2} \):

\[
\frac{s_{i}^{n+1/2}-s_{i}^{n-1/2}}{\Delta t}=f(u^{n})\Rightarrow \]

\[
s_{i}^{n+1/2}=s_{i}^{n-1/2}+\Delta tf(u^{n})\]
We could now write the staggered leapfrog method for equations like (\ref{leap-geral}):

\begin{equation}
\label{leapfrog-geral_discreto}
\begin{array}{c}
u_{i}^{n+1}=u_{i}^{n}+\Delta ts_{i}^{n+1/2}\\
s_{i}^{n+1/2}=s_{i}^{n-1/2}+\Delta tf(u^{n})
\end{array}
\end{equation}

It could be easily shown \cite{recipes} that this procedure is formally equivalent
to taking direct second-order finite differences for the second-order derivatives
in the wave equation, giving the following scheme:

\begin{equation}
\label{leapfrog_onda_1d}
\frac{u_{i+1}^{n}-2u_{i}^{n}+u_{i-1}^{n}}{\Delta x^{2}}=\frac{u_{i}^{n+1}-2u_{i}^{n}+u_{i}^{n-1}}{\Delta t^{2}}
\end{equation}

The situation with a dissipative term is a little bit more complicated. We said
above that the first derivatives in the staggered leapfrog method are to be
evaluated at the staggered points, but an equation with a dissipative term has
a first derivative evaluated at the real ones:

\begin{equation}
\label{leap-geral_dissip}
\frac{\partial s}{\partial t}=-\eta s+f[u(x,t)]\Rightarrow 
\end{equation}

\[
\frac{s_{i}^{n+1/2}-s_{i}^{n-1/2}}{\Delta t}=-\eta s_{i}^{n}+f[u_{i}^{n}]\]

Indeed, the above equation has an explicit first derivative evaluated at a real
point (\( s_{i}^{n} \)). How can we fix this problem? The trick is to take
an average over the adjacent points, so we recover the staggered points in the
dissipative term:

\[
\frac{s_{i}^{n+1/2}-s_{i}^{n-1/2}}{\Delta t}=-\eta \frac{s_{i}^{n+1/2}+s_{i}^{n-1/2}}{2}+f[u_{i}^{n}]\]

Continuing with the same procedure done for the previous case (we need an expression
for \( s_{i}^{n+1/2} \) to use with Eq. (\ref{leap-unp1})), we have:

\[
s_{i}^{n+1/2}+\frac{\eta \Delta t}{2}s_{i}^{n+1/2}=s_{i}^{n-1/2}-\frac{\eta \Delta t}{2}s_{i}^{n-1/2}+\Delta tf[u_{i}^{n}]\Rightarrow \]

\[
s_{i}^{n+1/2}=\frac{\left( 1-\frac{\eta \Delta t}{2}\right) s_{i}^{n-1/2}+\Delta tf[u_{i}^{n}]}{1+\frac{\eta \Delta t}{2}}\]

Now we have the staggered leapfrog scheme for equations like (\ref{leap-geral_dissip}):

\[
u_{i}^{n+1}=u_{i}^{n}+\Delta ts_{i}^{n+1/2}\]

\[
s_{i}^{n+1/2}=\frac{\left( 1-\frac{\eta \Delta t}{2}\right) s_{i}^{n-1/2}+\Delta tf[u_{i}^{n}]}{1+\frac{\eta \Delta t}{2}}\]

For the case of a wave equation in \( d \) dimensions, the \( f[u] \) term
will have a \( d \)-dimensional Laplacian \( \nabla ^{2}u(\vec{x},t) \), with
\( \vec{x}=(x_{1},x_{2},...,x_{d}) \), and this Laplacian is to be evaluated
numerically with a second-order finite difference scheme at the step \( n \):

\[
\nabla ^{2}u(\vec{x},t)\approx \frac{u_{i+1,j,k,...}^{n}-2u_{i,j,k,...}^{n}+u_{i-1,j,k,...}^{n}}{\Delta x_{1}}+\frac{u_{i,j+1,k,...}^{n}-2u_{i,j,k,...}^{n}+u_{i,j-1,k,...}^{n}}{\Delta x_{2}}+...\]

When solving the staggered leapfrog method for the first step \( n=1 \), we
need \( s_{i}^{1/2} \), but the initial conditions are usually defined at the
initial time \( n=1 \). One way to solve this problem is to integrate the first
step using the Euler scheme (which needs only the initial points at \( n=1 \))
with a time-step of \( \Delta t/2 \), thus obtaining the required value of
\( s_{i}^{1/2} \).

Now some words about the efficiency of this method. When we apply the von Neumann
stability analysis in leapfrog equations we arrive at the Courant condition
\cite{recipes}. When this condition is satisfied, the staggered leapfrog method
is not only stable, but also \emph{conservative}, in the sense that it does
not introduce any numerical dissipation.

For the applications which will be shown in Chapter \ref{sec: aplic}, we shall
use the following conditions:

\[
\left. u(t)\right| _{\mbox {boundaries}}=0,\: \: \: \: \left. \frac{\partial u(t)}{\partial t}\right| _{\mbox {boundaries}}=0\]

\[
\left. u(x,y)\right| _{t=0}=C\exp \left[ -\frac{(\vec{r}-\vec{r}_{0})^{2}}{2\gamma }\right] ,\]
where \( \vec{r}=x\hat{i}+y\hat{j} \), \( \vec{r}_{0}=\frac{l}{2}\hat{i}+\frac{l}{2}\hat{j} \),
\( l \) is the lattice length, \( C \) is a normalization constant, and \( \gamma  \)
is a sufficiently small constant such that \( u(\vec{r})\rightarrow 0 \) as
\( \vec{r}\rightarrow  \)boundaries, that is, the initial condition is a gaussian
sufficiently localized to make \( u \) smooth at the boundaries.

\chapter{Examples\label{sec: aplic}}

\section{The Free String (1D)}

These simulations were executed in a PC of \( 350MHz \), for lattices of at
most \( N=1000 \). The integration time lay in the order of seconds. 

\begin{figure}
{\centering \resizebox*{0.5\textwidth}{!}{\includegraphics{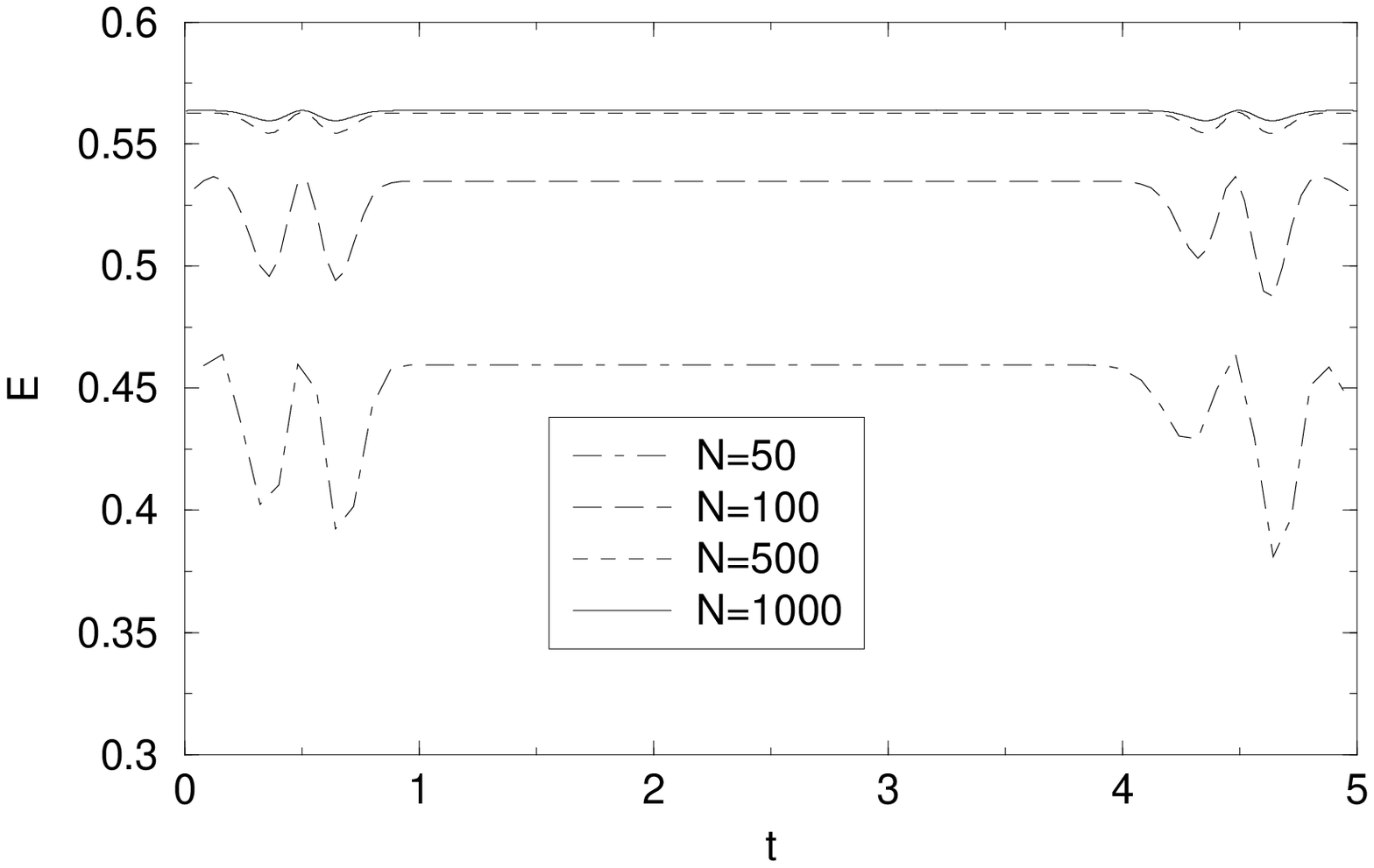}} 
\resizebox*{0.5\textwidth}{!}{\includegraphics{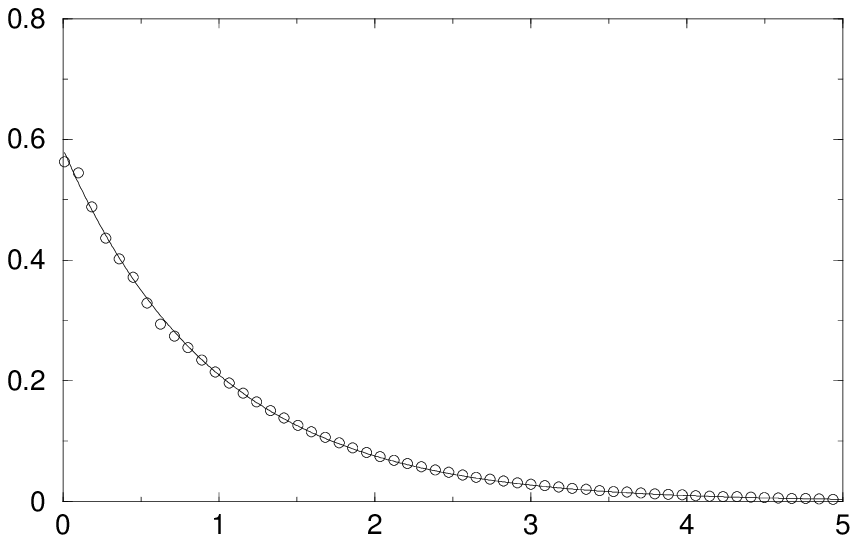}} \par}

\caption{\label{fig: et_para_ns}\protect\( E\times t\protect \) for different lattice
spacings and \protect\( \eta =0\protect \) (above) and \protect\( E\times t\protect \)
for \protect\( N=1000\protect \) and \protect\( \eta =1\protect \) (below).
The linear dimension is fixed at \protect\( L=1\protect \).}
\end{figure}\begin{figure}
{\centering \resizebox*{0.5\textwidth}{!}{\includegraphics{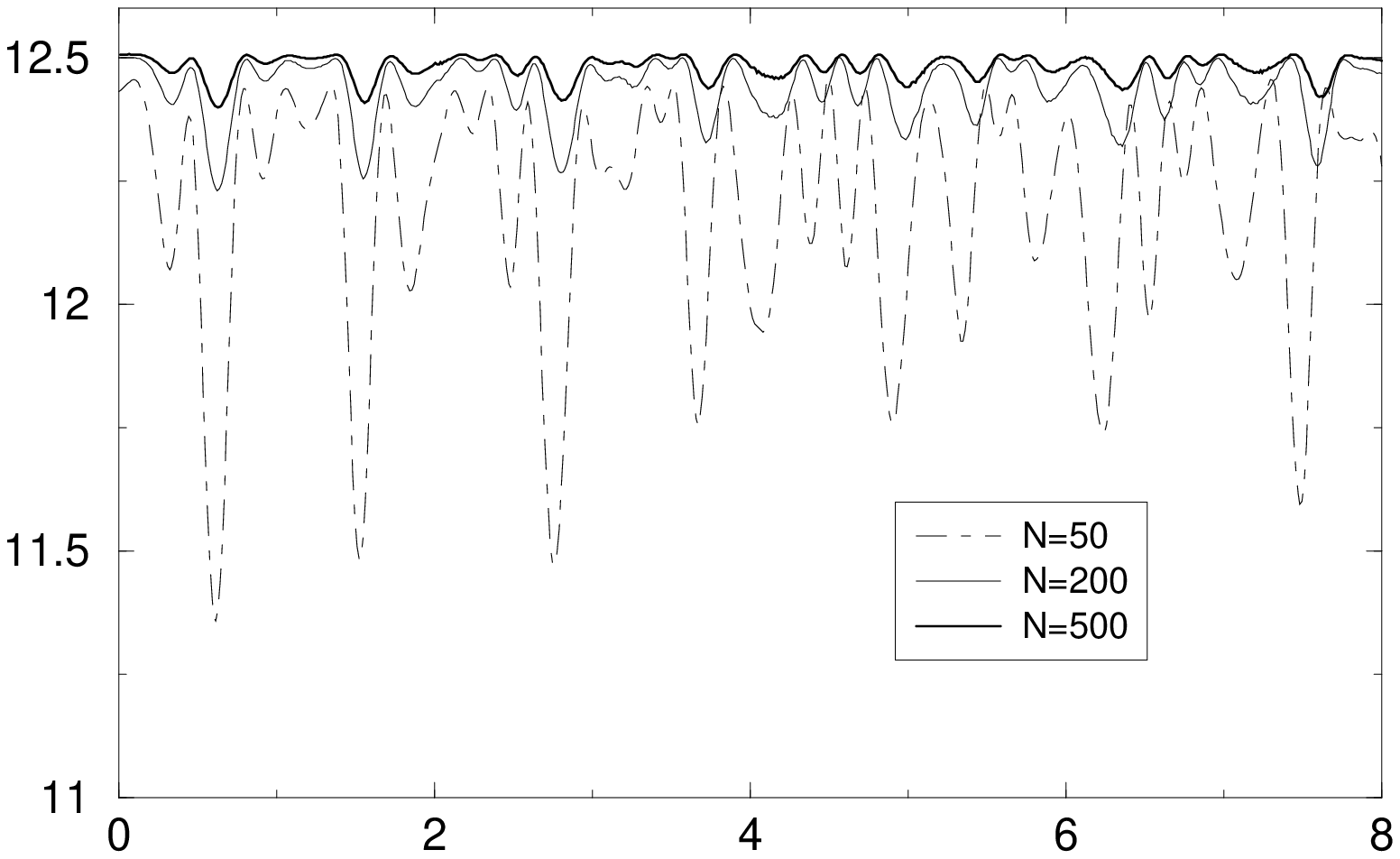}} \resizebox*{0.5\textwidth}{!}{\rotatebox{-90}{\includegraphics{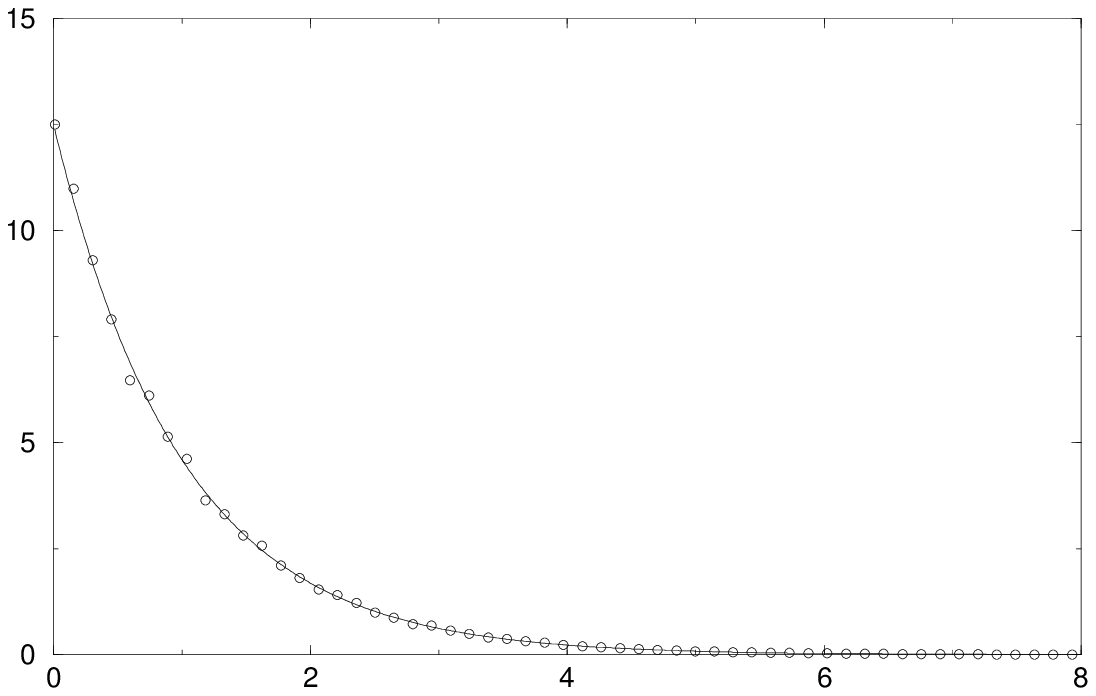}}} \par}

\caption{\label{fig: memb_eta0}\protect\( E\times t\protect \) for \protect\( \eta =0\protect \)
and various \protect\( N\protect \) (above) and \protect\( E\times t\protect \)
for \protect\( \eta =1\protect \) and \protect\( N=200\protect \) (below).
The dimensions of the membrane are \protect\( L\times L=1\protect \) for all
runs.}
\end{figure}

Figure \ref{fig: et_para_ns} shows some results for the conditions of the previous
section for various parameters. Notice the improvement of energy conservation
for finer resolutions and the very good exponential fitting for \( \eta =1 \).
This exponential result is expected, since the vibrating string could be understood
in terms of the Fourier space, where each mode behaves as a decoupled harmonic
oscillator with a damping given by \( \eta  \).

\section{The Membrane (2D)}

These simulations were also executed in a PC of \( 350MHz \), and for lattices
of \( N=500 \) the integration time reached half an hour.

Figure \ref{fig: memb_eta0} shows some results for \( \eta =0 \) and \( \eta =1 \).
For a non-conservative system (\( \eta =1 \)), we see also the exponential
fitting for \( N=200 \).

\appendix

\chapter{Taylor's Theorem\label{sec: Ap_taylor}}

\section{Definitions}

When we want to transform a differential equation into a difference equation,
the Taylor expansion is often used:
\begin{eqnarray}
f(x) & = & f(x_{0})+f'(x_{0})(x-x_{0})+\frac{1}{2}f''(x_{0})(x-x_{0})^{2}+...=\\
 & = & \sum ^{\infty }_{n=0}\frac{f^{(n)}(x_{0})}{n!}(x-x_{0})^{n},\label{Taylor_serie} 
\end{eqnarray}
where \( x_{0} \) is the point around which we want to expand \( f(x) \)\footnote{
For \( x_{0}=0 \) this expansion is also known as \emph{Maclaurin expansion}.
}. An alternative form for this expansion is achieved doing a simple variable
change \( x\rightarrow x+h \) and \( x_{0}\rightarrow x \):
\begin{equation}
\label{Taylor_serie_h}
f(x+h)=\sum _{n=0}^{\infty }\frac{f^{(n)}(x)}{n!}h^{n}
\end{equation}

In the numerical case we will be interested in \emph{truncating} the series,
so that we finish with a finite number of terms. We could then write this expansion
in the following form:

\begin{equation}
\label{Taylor_trunc}
f(x+h)=\sum _{n=0}^{m}\frac{f^{(n)}(x)}{n!}h^{n}+{\cal O}(h^{m+1}),
\end{equation}
where \( {\cal O}(h^{m+1}) \) corresponds to the truncated terms which powers
of \( h \) are equal or higher than \( m+1 \) (this term is frequently called
the \emph{error of order \( m+1 \)}). Notice that under the numerical point
of view it is important to know the order of \( {\cal O} \) in the discretization,
since for \( h\ll 1 \), the greater the order of \( {\cal O} \) the more negligible
will be the error.

\section{Useful Expansions}

Some expansions that will be used throughout this text are shown below. All
of them could be obtained from (\ref{Taylor_trunc}) by directly solving for
the desired term or using more than one expansion to find higher order expansions
for the derivative, and then solving the system. For instance:
\[
\left\{ \begin{array}{cc}
f(x+h) & =f(x)+f'(x)h+\frac{1}{2}f''(x)h^{2}+...\\
f(x-h) & =f(x)-f'(x)h+\frac{1}{2}f''(x)h^{2}-...
\end{array}\right. \]

\begin{equation}
\label{ap_flinha_1a_ordem}
f'(x)=\frac{f(x+h)-f(x)}{h}-{\cal O}(h)
\end{equation}

\begin{equation}
\label{ap_flinha_2a_ordem}
f'(x)=\frac{f(x+h)-f(x-h)}{2h}-2{\cal O}(h^{2})
\end{equation}

\begin{equation}
\label{ap_f2linha_2a_ordem}
f''(x)=\frac{f(x+h)-2f(x)+f(x-h)}{h^{2}}-2{\cal O}(h^{2})
\end{equation}

\begin{equation}
\label{ap_dfdx_1a_ordem}
\frac{\partial f(x,y)}{\partial x}=\frac{f(x+h,y)-f(x,y)}{h}-{\cal O}(h)
\end{equation}

\begin{equation}
\label{ap_d2fdx2_2a_ordem}
\frac{\partial ^{2}f(x,y)}{\partial x^{2}}=\frac{f(x+h,y)-2f(x,y)+f(x-h,y)}{h^{2}}-2{\cal O}(h^{2})
\end{equation}
Notice that when we divide an error of order \( {\cal O}(h^{n}) \) by \( h^{r} \),
automatically this error turns to order \( n-r \), i.e., \( {\cal O}(h^{n})/h^{r}={\cal O}(h^{n-r}) \).

\end{document}